# Multistate Tuning of Third Harmonic Generation in Fano-Resonant Hybrid Dielectric Metasurfaces

Omar A. M. Abdelraouf [1,3], Aravind P. Anthur[3], Zhaogang Dong[3], Hailong Liu[3], Qian Wang[3], Leonid Krivitsky[3], Xiao Renshaw Wang[1,2]*, Qi Jie Wang[1,2,]*, Hong Liu[3,]*

**Affiliations:**
[1]School of Physical and Mathematical Sciences, Nanyang Technological University, Singapore 637371, Singapore
[2]School of Electrical and Electronic Engineering, 50 Nanyang Avenue, Nanyang Technological University, Singapore 639798, Singapore
[3]Institute of Materials Research and Engineering, Agency for Science, Technology and Research (A*STAR), 2 Fusionopolis Way, #08-03, Innovis, Singapore 138634, Singapore

*To Whom correspondence should be addressed. Email: renshaw@ntu.edu.sg, qjwang@ntu.edu.sg, h-liu@imre.a-star.edu.sg

**ABSTRACT**

Hybrid dielectric metasurfaces have emerged as a promising approach to enhancing near field confinement and thus achieving high optical nonlinearity using low loss dielectrics. Additional flexibility in design and fabrication of hybrid metasurfaces allows dynamic control of light, which is value-added for a wider range of applications. Here, we demonstrate a tunable and efficient third harmonic generation (THG) via hybrid metasurfaces with phase change material $Ge_2Sb_2Te_5$ (GST) deposited on top of amorphous silicon nanostructutes. Fano resonance is excited to confine the incident light inside the hybrid metasurfaces, and an experimental quality factor (*Q*-factor) of 125 is achieved at the fundamental pump wavelength around 1210 nm. We demonstrate the switching between a turn-on state of Fano resonance in the amorphous state of GST and a turn-off state in its crystalline state and also gradual multistate tuning of THG emission at its intermediate state. We achieve a high THG conversion efficiency of $\eta = 2.9 \times 10^{-6}$ %, which is more than ~32 times of that of a GST-based Fabry-Pèrot cavity under a similar pump laser power, thanks to the enhanced field confinement due to the Fano resonance. Our results show the strong potential of GST-based hybrid dielectric metasurfaces for efficient and tunable nonlinear optical devices.

# INTRODUCTION

Third harmonic generation (THG) is a common nonlinear optical phenomenon,[1] which generates photons with triple the frequencies of those of the incident photons. Unlike a second harmonic generation (SHG) the THG can be generated from nonlinear materials with arbitrary lattice structures. THG has been used in many applications, such as spectroscopy,[2] imaging of lipid bodies in tissues,[3] generating free-electron laser in the near infrared (NIR),[4] probing quantum tunneling limit in self-assembled monolayer,[5] and light modulation.[6] Furthermore, its tunability has been explored in light modulation of generated harmonics.[7] Motivated by those applications, it is highly desirable to improve the conversion efficiency of THG and to be able to actively modulate its emission amplitude. Nonlinear crystals have been widely utilized for THG. However, they are bulky and have stringent phase-matching conditions between the fundamental and the generated harmonic fields. Furthermore, they need an external cavity to enhance the THG efficiency. On the other hand, metasurfaces, namely artificially-engineered nanostructures, have been utilized to confine the incident light down to a subwavelength scale to enhance THG. In addition, metasurfaces do not require stringent phase-matching conditions for high-order harmonic generation (HHG). Plasmonic metasurfaces consisting of various metallic nanostructures have been used for THG such as gap plasmon,[8] multi-quantum well,[9] and toroidal dipole plasmonic meta-atom.[10] Generally, the light confinement via plasmonic metasurfaces is constrained to only over the surface of metals. Moreover, the intrinsic Ohmic losses of metals reduce both conversion efficiency and the damage threshold of plasmonic metasurfaces.

In contrast, owing to their high permittivity and inherent low optical losses, dielectric metasurfaces have been proposed to overcome the issue of Ohmic losses of metallic metasurfaces.[11,12,13] The modes of the incident light are confined in the whole volume of dielectric nanostructures, which greatly enhance the conversion efficiency of THG. In addition, dielectric metasurfaces have higher damage thresholds than the plasmonic counterparts. Recently, dielectric metasurfaces have been extensively studied to enhance the efficiency of THG and other HHGs in nonlinear dielectric nano-resonators driven by high Q-factor Fano-resonance,[14,15,16,17] anapole resonance,[18,19,20,21] and bound states in the continuum (BIC).[22,23,24] The reconfigurable metasurfaces have been developed to realize active tuning of various functions upon different stimuli.[25,26] For example, tuning can be achieved via changing the refractive index of liquid crystals by manipulating their molecules rotation,[27] using optical pumping to tune free carriers generation in semiconductors,[28] or applying an electrical bias to tune Fermi energy level in graphene.[29]

Integrating phase change materials (PCMs) in hybrid dielectric metasurfaces offers reversible phase change between amorphous and crystalline states, ultrafast switching speeds, broadband and large turning of the refractive index, and process compatibility with the CMOS technology.[30,31,32] A Fabry-Pèrot cavity in reflection mode based on high refractive index $Ge_2Sb_2Te_2$ (GST) has been demonstrated to achieve two-state switching (ON/OFF) of THG.[33] However, its THG conversion efficiency was only about $5.46 \times 10^{-9}$%, whihis attributed mainly to the low Q-factor of the device and inherent high optical losses in the thick GST film.

In this work, we design transmissive mode hybrid metasurfaces consisting of amorphous silicon and GST nanostructures, and experimentally realized gradual tuning of the amplitude of THG. The high nonlinearity of silicon nanostructures generates third harmonic emission. At the same time, the gradual tunability of THG amplitude is realized via the phase change among the amorphous, intermediate and crystalline states of the GST nanostructures.[34,35] Fano resonance subsequently enhanced the light confinement inside the silicon nanostructures, significantly improving the THG conversion efficiency.

**RESULTS AND DISCUSSION**
Figure 1 illustrates the design concept of tunable THG using hybrid silicon-GST metasurfaces, consisting of two split-ring resonators (SRR) connected to each other as shown in Fig. 1a. The width of the middle rod ($w_1$) and the outer rod width ($w_2$) are optimized to be 200 nm and 120 nm, respectively, to maximize the Q-factor of Fano resonance (see Fig. S1 in the Supplementary Information (SI)). The silicon SRR structure enhances the field confinement and hence the Q-factor of the Fano resonance, which is induced by breaking the symmetry of SRR of the hybrid metasurfaces. The incident light is perpendicular to the sample surface (x-y direction), and its polarization is along the y-axis to excite the Fano resonance. Other structural parameters of SRR on quartz are shown in Fig. 1b. A 5-nm thick GST layer on top of the silicon SRR enables both tunable THG in transmission mode and tunable Fano resonance in the silicon SRR. The nanofabrication process of the metasurfaces started from electron beam lithography (EBL) followed by inductively coupled plasma-reactive ion etching (ICP-RIE) using hydrogen silsesquioxane (HSQ) as an etching mask. A scanning electron microscope (SEM) image of the fabricated hybrid metasurfaces is shown in Fig. 1c, and its inset shows the SEM image of the unit cell. For more details of the fabrication steps, see the Methods section and Fig. S3 in the SI.

The phase change of GST can be excited upon annealing of samples on a hotplate at a temperature above its phase-change temperature. Baking GST film at ~ 180 °C and ~ 300 °C for a few minutes will transform it into an intermediate state and crystalline state, respectively. Intermediate state of GST film refers to the metastable phase (the cubic phase) while the crystalline state denotes the co-existence of the cubic phase and the hexagonal phase with different ratios, or a pure hexagonal phase (stable phase).[31] We investigated optical properties of GST at the multiple phase states by measuring the refractive index ($n$) and extinction coefficient ($k$) of GST film using variable angle spectroscopic ellipsometry (JA Woollam VB400). The fitting was carried out by using the Tauc-Lorentz model and the mean square error (MSE) of less than 10 was achieved for all measurements. The measured refractive index and extinction coefficient are plotted in Figs. 1d and 1e, respectively. It can be observed that the refractive index gradually increased with the annealing time. At the wavelength of ~ 1200 nm, the changes in refractive index (Δn) of ~ 0.68 and 0.79 were measured after annealing at a temperature of $180^{0}$C for 2 minutes and 4 minutes, respectively. Similarly, the extinction coefficient gradually increased from an amorphous state (a-GST) to an intermediate state (i-GST) and finally reached the maximum value at the crystalline state (c-GST). The corresponding changes of extinction coefficient (Δ$k$) were ~ 0.50 and 0.61, respectively. Multiple i-GSTs can be achieved by increasing the annealing time. Upon c-GST, a very large value of Δ$n$ of ~ 3.15 and corresponding Δ$k$ of ~ 2.37 at the wavelength of 1200 nm were measured respectively, which agreed well with the literature.[36] To validate the quality of c-GST film (5 nm) sputtered on quartz, we did Raman spectroscopy measurement and the measured spectra of a-GST and c-GST films (see Fig. S5) agreed with the literature.[37]

Figure 2 shows the simulated and measured results verifying the multistate tunability of Fano-resonance in the hybrid dielectric metasurfaces. As shown in Fig. 2a, a sharp peak of Fano resonance is formed at the wavelength of 1292.6 nm in the simulated transmission spectrum of a-GST. Due to the low optical loss at a-GST, it leads to the strongest Fano-resonance and the highest Q-factor. As a result, high transmission of ~ 72% at the Fano resonance wavelength is achieved, which denotes the "turn-on" state of the device. Fano resonance provides a high Q-factor of ~615 that corresponds to a full width at half maximum (FWHM) of around 2.1 nm. The dip depth of Fano resonance is around 38%, which was measured through the amplitude difference between the maximum and the minimum transmission at the Fano resonance wavelength. For the intermediate state achieved after annealing the sample at 180°C for 2 minutes, the dip depth reduces to 28%. The FWHM of Fano resonance increases to 2.4 nm while the Fano resonance wavelength redshifts to 1292.9 nm. After 4 minutes of annealing, the dip depth is further reduced to 8%. The FWHM of Fano resonance increases to 5.6 nm while the resonance wavelength redshifts to 1294.6 nm. At i-GST, the FWHM of Fano resonance increases with the annealing time, which shows the Fano-resonance becomes weaker due to the increase of the optical loss in GST film. The i-GST denotes the "dynamic tuning" state of the device since its optical properties gradually vary with the phase change upon annealing. The crystalline state is achieved by annealing the sample at 300°C for 5 minutes[38], and there is no resonance observed for c-GST. The highest optical loss in c-GST state does not excite the Fano resonance so that the hybrid metasurfaces cannot function for the THG enhancement, which denotes the "turn-off" state of the device. All simulations were carried out using the measured refractive index of amorphous silicon and GST at different states. The measured transmission spectrum is shown in Fig. 2b. The curve of a-GST exhibits Fano resonance wavelength at 1211.8 nm, FWHM of 9.7 nm and dip depth of 24%, respectively. The measured Q-factor was ~125, which is lower than the simulated value, possibly due to inaccuracy in the device fabrication. In particular, a slight variation of sidewall width significantly blueshifts the Fano resonance wavelength. At i-GST, the dip depth of Fano resonance gradually reduced from 17% to 7% when the annealing time rose from 2 mins to 4 mins. At the same time, it was observed that the FWHM of the Fano resonance curve increased from ~10.3 nm to 12.5 nm correspondingly. The Fano resonance completely vanished at c-GST state, which is identical to the theoretical prediction. The dip depth of the Fano resonance gradually reduced from a-GST to i-GST till it became zero at the c-GST with a measured transmission of ~55%.

To investigate the origin of Fano-resonance, the multipolar decomposition method has been implemented individually using the finite difference time domain (FDTD) method for each GST state and the simulated results are shown in the top row of Fig. 2c.

- For a-GST, the total scattering cross-section has a strong electric quadrupole (*EQ*) resonance with a FWHM of ~ 2.4 nm, while the electric dipole (*ED*) shows damped oscillation with a minimal normalized magnitude of less than 0.028. Therefore, the Fano resonance in the proposed hybrid metasurfaces arises from destructive interference between the in-plane damped electric dipole oscillating in the *y*-direction ($ED_y$) and the out-of-plane high Q-factor electric quadrupole in *z*-direction ($EQ_z$).
- For i-GST (after 2 minutes of annealing), the normalized scattering cross-section shows a larger FWHM (3.2 nm) of $EQ_z$ curve than that of a-GST. Meanwhile, the peak $EQ_z$ scattering degrades by 33%. $ED_y$ maintains a small scattering magnitude of less than 0.015. That explains the origin of relatively weak Fano-resonance at i-GST.
- At c-GST, the total normalized scattering intensity has been reduced significantly to ~0.12 as compared with a-GST. No resonance is observed for all modes, resulting in switching off the Fano resonance, which agrees well with the measured transmission of the metasurfaces.

The simulated results of the normalized total electric field intensity of a-GST, i-GST, and c-GST are shown in the bottom row of Fig. 2c. In each figure, the normalized electric field profile of *x*-*y* (horizontal) plane is at the position of *z*=200 nm.

- For a-GST, the field distribution shows high field confinement in the *EQ* mode with out-of-plane field oscillation. In contrast, the *ED* mode oscillates horizontally along the central air gap.
- For i-GST , the corresponding normalized field distribution shows the field vectors of *EQ* and *ED* similar to a-GST. However, the intensity of field confinement of i-GST is reduced to 67% of a-GST due to the higher optical losses in i-GST film, which results in a weaker Fano resonance.

- For c-GST, the field distribution shows the vanishing of *EQ*, indicating there is no out-of-plane field oscillation. Meanwhile, the maximum field intensity is decreased to 29% compared with a-GST. The field distribution and the multipolar scattering agree with each other in terms of vanishing resonant modes near Fano resonance wavelength at a-GST.

To experimentally measure the near field enhancement achieved by Fano resonance, we used a home-built optical setup. The femtosecond laser beam is generated from an optical parametric oscillator (APE) with a pulse width of 200 fs and a tunable range of 1100 ~ 1300 nm, which is schematically shown in Fig. 3a. The incident light is focused on the sample using a plano-convex lens with a focal length of 25.4 mm. The diameter of the focused laser beam spot is 15 μm. The transmitted THG is collected by an objective with 0.55 numerical aperture (NA) and filtered by low pass UV filter before being coupled to a spectrometer (Ocean Optics, USB4000). Details of the optical setup can be found in the Methods section and Fig. S4 in the SI.

Fig. 3b shows the measured optical characterization of THG at a-GST, which demonstrates a THG peak at the wavelength of ~404 nm, corresponding to the fundamental incident wavelength of ~1216 nm. It indicates that the enhancement of THG emission comes from the enhanced localized near field inside hybrid metasurfaces at the Fano resonance wavelength, as shown in Fig. 2b. Moreover, the emission peak wavelength is nearly one-third of the Fano resonance peak wavelength, indicating that the emitted field is THG. We further examined the polarization dependence of THG emission by varying the pump laser polarization via a half-wave plate (HWP). Fig. 3c plots the variation of the measured intensity of THG emission with the pumped laser polarization. The enhancement of THG emission vanishes under the perpendicular incident polarization (*x*-direction). It is consistent with our design as depicted in Fig. 1 that Fano resonance is only able to be excited by the *y*-polarized pump laser. The asymmetrical distribution of intensity in the polar plot (Fig. 3c) is mainly due to the imperfect alignment between the fabricated sample and the polarizer. The far field emission of THG was measured using a 4f-lens setup to obtain a back focal plane (BFP) image using CCD camera. The measured diffraction patterns are plotted in Fig. 3d. The integration time used in the measured BFP image was calibrated to avoid color saturation. The detected intensity of the higher-order diffraction is lower than that of the zero-order diffraction, which indicates the feasibility of collecting most of THG emission from the developed hybrid metasurfaces using low NA lenses.

The measurement and characterization results of the tunability of THG are plotted in Fig. 4. We measured the THG spectrum at each state of GST and fixed the output power of the laser to the maximum of ~55 mW, which was equivalent to a peak power density of up to 1.6 GW/cm$^2$ as shown in Fig. 4a. At a-GST, the intensity of THG emission reached its maximum at the peak wavelength of ~ 404 nm. After annealing for 2 minutes, the measured THG emission was degraded. That could be explained by the weaker Fano resonance after annealing, as shown in Fig. 2b. When the annealing time increased to 4 minutes, the emitted THG further decreased due to the poor light confinement of the even weaker Fano resonance. For c-GST, no THG emission was detected, which indicated that thin GST film had undergone a complete phase change, and there was no observation of Fano resonance. The modulation depth (MD) of the developed hybrid metasurface can be calculated through

$$MD = \frac{|I_{max} - I_{min}|}{I_{max}} \qquad (1)$$

where $I_{max}$ and $I_{min}$ are the maximum and the minimum THG intensity. By excluding the background noise of the THG signal, we achieved MD of ~65 % for i-GST (2 min), ~78% for i-GST (4 min), and ~100% for c-GST. The measured MD results show the potential for efficient modulation of THG intensity.

We also measured the THG power dependence on the pump power at a-GST state. Fig. 4b shows the variation of the measured THG power with pump powers. We achieved the highest THG power of up to 1.6 nW at the pump power of 55 mW at the a-GST. That corresponds to a THG enhancement factor of ~125 compared with an unpatterned GST-Si film. Our THG conversion efficiency was calculated based on the following equation:

$$\eta = \frac{P_{THG}}{P_{pump}} = 2.9 \times 10^{-6} \,\% \qquad (2)$$

The experimentally measured THG conversion efficiency is ~32 times higher than the amplitude of THG emitted from the Fabry-Perot cavity incorporated with GST using the same pump laser power.[33] To fit the measured THG power, we used the power dependence formula $P_{THG} = aP^b_{Pump}$ with a line slope of ~3. It fits well the THG power for a-GST state with R-squared values higher than 0.995, which indicates that the measured power was dominantly generated by THG.

**CONCLUSION**

In conclusion, we have demonstrated multistate tuning in THG emission using hybrid metasurfaces based on silicon-GST metasurfaces. We used silicon as a high dielectric constant and low loss material at the near-infrared spectrum to enable Fano resonance and experimentally achieve a Q-factor of up to 125 for THG emission enhancement. A thin layer of 5 nm thick phase change material of GST was deposited on top of Si metasurfaces to gradually tune the THG emission in the transmission mode. THG conversion efficiency up to $2.9 \times 10^{-6}$ % was achieved, which is ~32 times higher than a recent work based on GST-based Fabry-Pèrot cavity. Moreover, we have demonstrated multiple intermediate states of GST, which results in multilevel emission amplitude of THG. We believe that our devicess pave the way for tunable nonlinear optics applications such as light modulation of HHGs in nonlinear materials, tunable lasing, and quantum entanglement.

**METHODS**

**Transmission and Field Profile Simulations**

Simulation of the proposed tunable silicon-GST hybrid dielectric metasurfaces was carried out using a finite difference time domain program (Lumerical FDTD Solutions) [39]. Three-dimensional simulations with periodic boundary conditions in the lateral direction (*x-y* plane) and perfectly matched layer in the *z*-direction were adopted. The incident polarization was parallel to the middle rod of the structure (*y*-direction). The maximum mesh size was 10 nm for all simulation domains except phase change material of GST, which was smaller than 0.1 nm. The refractive index of amorphous silicon and GST was measured from their thin films deposited on a silicon substrate via ellipsometer, and the results are plotted in Fig. S4 in the SI and Figs. 1d-1e in the main text, respectively. The refractive index of quartz was imported from Lumerical's database.

**Multipolar Decomposition Simulations**

The multipolar decomposition was simulated using two three-dimensional monitors in Lumerical FDTD, i.e. one for refractive index and the other for getting field components at each mesh. The scattering cross-sections of electric and magnetic modes were calculated using the following equations.

$$C_{ED} = \frac{k_0^4}{6\pi\epsilon_0^2 E_0^2}\left|P_{car} + \frac{ik_0}{c}\left(t + \frac{k_0^2}{10}\overline{R_t^2}\right)\right|^2 \tag{1}$$

$$C_{MD} = \frac{\eta_0^2 k_0^4}{6\pi E_0^2}\left|m_{car} - k_0^2\overline{R_m^2}\right|^2 \tag{2}$$

$$C_{EQ} = \frac{k_0^6}{80\pi\epsilon_0^2 E_0^2}\left|\overline{\overline{Q_e}} + \frac{ik_0}{c}\overline{\overline{Q_t}}\right|^2 \tag{3}$$

$$C_{MQ} = \frac{\eta_0^2 k_0^6}{80\pi E_0^2}\left|\overline{\overline{Q_m}}\right|^2 \tag{4}$$

Where $C_{ED}$, $C_{MD}$, $C_{EQ}$, $C_{MQ}$ refer to scattering cross section of an electric dipole, magnetic dipole, electric quadrupole, and magnetic quadrupole respectively. The scattering power formulas of electric dipole ($P_{car}$), toroidal dipole ($t$), magnetic dipole ($m_{car}$), electric quadrupole ($\overline{\overline{Q_e}}$), and magnetic quadrupole ($\overline{\overline{Q_m}}$) can be found in the literature [40].

**Fabrication of hybrid metasurfaces**

Figure S3 schematically shows the fabrication flow process. Initially, the quartz substrate from (Photonik Singapore) was cleaned using acetone and isopropanol alcohol (IPA) in an ultrasonic bath for 5 min. A layer of ~400 nm thick amorphous silicon was deposited using inductively coupled plasma chemical vapor deposition (ICP-CVD, Oxford Plasmalab System 380) at a substrate temperature of 250 °C with RF power of 50 W, ICP power of 3000 W, silane and argon gases with the flow of 45 and 30 sccm, respectively. 5 nm of phase change material of GST was deposited via an unbalanced magnetron sputtering under a power of 30 W, argon gas flow of 20 sccm, and a chamber pressure of 4 mTorr. Hydrogen silsesquioxane (HSQ) was used as an etching mask for patterning the silicon metasurfaces using electron beam lithography (EBL, Elionix ELS-7000). SurPass 3000 was used as an adhesion promoter followed by a spin coating of HSQ (6%) at a speed of 2000 rpm for 60 sec to get the thickness of around 100nm. After that, EZspacer was spin-coated at 1500 rpm for 30 sec to avoid charging effect during EBL exposure, followed by nitrogen drying to remove excess EZspacer. EBL exposure was carried out under the conditions of the electron beam current of 500 pA, acceleration voltage of 100 kV, and dose charge 9600 uc/cm² to expose a field of 300 x 300 um² with 60,000 dots. After exposure, the sample was rinsed with deionized (DI) water to remove EZspacer layer, then developed in salty solution NaOH/NaCl (1:4) for 60 sec, followed by another 60 sec to rinse in DI water. After that sample was rinsed using IPA and then dried with nitrogen gas. Etching of silicon metasurfaces was implemented using inductively coupled plasma reactive ion etching (ICP-RIE, Oxford OIPT Plasmalab system) with HSQ as an etching mask. Etching gas consisted of chlorine with a flow of 22 sccm, using DC power of 200 W and ICP power of 400 W at room temperature. All measurements were done with a top residual HSQ layer of thickness ~ 40nm to protect GST from oxidation during annealing process.

**THG measurement**

The optical setup used in this work is shown in Fig. 3a. Optical parametric oscillator (OPO) laser pumped via a Ti-Sapphire laser of wavelength 830 nm, with a repetition rate of around 76 MHz and pulse width of around 200 fs. The wavelength of OPO laser is tunable from 1100 nm to 1300 nm. Switching between OPO laser and ns-laser was done with a flipping mirror (M1). A variable attenuator was used to examine the power slope of THG signal by changing the pump laser power. A quarter-wave plate (QWP) was used to convert the elliptically polarized light to the linearly polarized light with a phase shift $\gamma$ from the horizontal plane. A half-wave plate (HWP), attached to a motorized stage, was used to control the linearly polarized light to get the maximum power after passing through a linear polarizer (LP), which was used for measuring the polarization dependence of THG. A dichroic mirror was used to cut off light with a wavelength shorter than 700nm. A lens (L1) of a focal length of 25.4 mm focused the laser beam on the sample. White light LED was used to align a laser spot on the fabricated pattern. The transmitted light was collected using an objective of 50x magnification and 0.55 numerical aperture (Mitutoyo Plan Apo). The collimated light was directed to a low pass filter (NE03A-A) with a cut-off wavelength above 400 nm. Flipping mirror (M2) switched among the back focal plane imaging of THG, THG spectrum, and metasurfaces transmission at NIR. A set of mirrors (L2 to L4) was used for back focal plane imaging. A spectrometer (Ocean Optics, USB4000) was used for the measurement of THG spectrum. For the diffraction measurement of THG signal, an objective x100 with numerical aperture 0.95 (Nikon Plan Apo) was used to collect all diffracted THG signal.

**Material characterization**

The measured refractive index of amorphous silicon was fitted using Tauc-Lorentz model. Mean square error (MSE) for the fitted data was lower than 10; the measured data was plotted in Fig. S4. To verify the phase change of a thin layer of GST (~5 nm, after annealing at a temperature of 300 $^0$C for 5 minutes), we carried out the measurement of Raman spectra of both a-GST and c-GST, and the results are plotted in Fig. S5a and S5b, respectively. A broad Raman scattering peak for a-GST centered near the frequency of 140 cm$^{-1}$, while c-GST film shows a narrow peak centered near the frequency of 117 cm$^{-1}$. It is consistent with the literature,[37] which validates the phase change of the thin GST layer.

**ASSOCIATED CONTENT**
**SUPPORTING INFORMATION**
Additional simulation results of tuning Fano resonance with different width of middle rod of the unit cell. The electric field maps inside GST film for amorphous, intermediate, and crystalline states. Fabrication flow steps of hybrid metasurfaces. The measured refractive index of amorphous silicon used in simulation. Measured Raman spectrum of GST film with 5 nm thickness for amorphous and crystalline states.

**AUTHOR CONTRIBUTIONS**
O. A. M. A. proposed structure design, performed simulations, materials characterization, and nanofabrication. Optical characterizations and measurements were performed by O. A. M. A. with the help of A. P. A. and L. K., who designed and built the THG setup. O. A. M. A. drafts the manuscript. All authors discussed and commented on the results and the manuscript. X. R. W, Q. J. W., and L. H. conceived the idea, supervised the project, and finalized the manuscript.

**Competing Financial Interests**
The authors declare no competing financial interest.

**ACKNOWLOEDGMENT**
This work is partially supported by Singapore Ministry of Education Academic Research Fund Tier 2 under grant no. MOE2018-T2-1-176, MOE-T2EP50120-006 and by A*STAR AME programmatic grant (grant no. A18A7b0058), IMRE project (SC25/18-8R1804-PRJ8), AME IRG grant (Project No. A20E5c0094 and A20E5c0095), and A*STAR Quantum Technologies for Engineering (QTE) grant No. A1685b0005. Authors would like to thank DENG Jie, YAP Sherry, HUANG Aihong, and Febiana Tjiptoharsono for their technical guide during the optimization of nanofabrication processes. Also, Ramon Paniagua-Dominguez, and Ho Jin Fa for their comments on multipolar decomposition results.

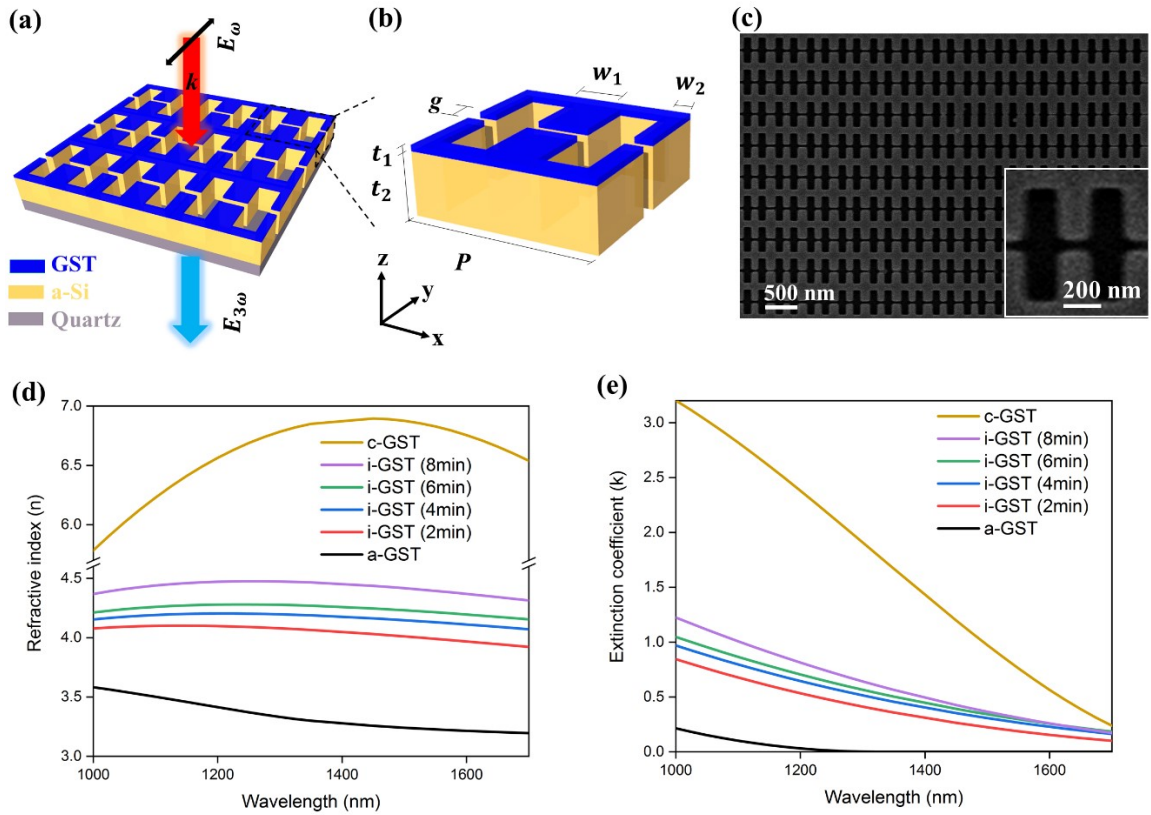

**Figure 1.** Schematics of the design of tunable THG via the hybrid silicon-GST metasurfaces. (a) Schematic of the proposed hybrid metasurfaces. The pump fs-laser ($E_\omega$) and transmitted THG ($E_{3\omega}$) signals are denoted by the red and blue arrows, respectively, where $\omega$ is the fundamental frequency. (b) Schematic of the hybrid metasurfaces unit cell with period $p$ = 700 nm, GST thickness $t_1$ = 5 nm, silicon thickness $t_2$ = 400 nm, middle air gap $g$ = 50 nm, middle rod width $w_1$ = 200 nm, and outer rod width $w_2$ = 120 nm. (c) SEM of the fabricated hybrid metasurfaces, and the inset shows the unit cell. (d) Measured real refractive index *n* for amorphous and crystalline GST, and different intermediate GST states after different annealing time at 180 $^0$C. (g) Corresponding measured extinction coefficient *k*.

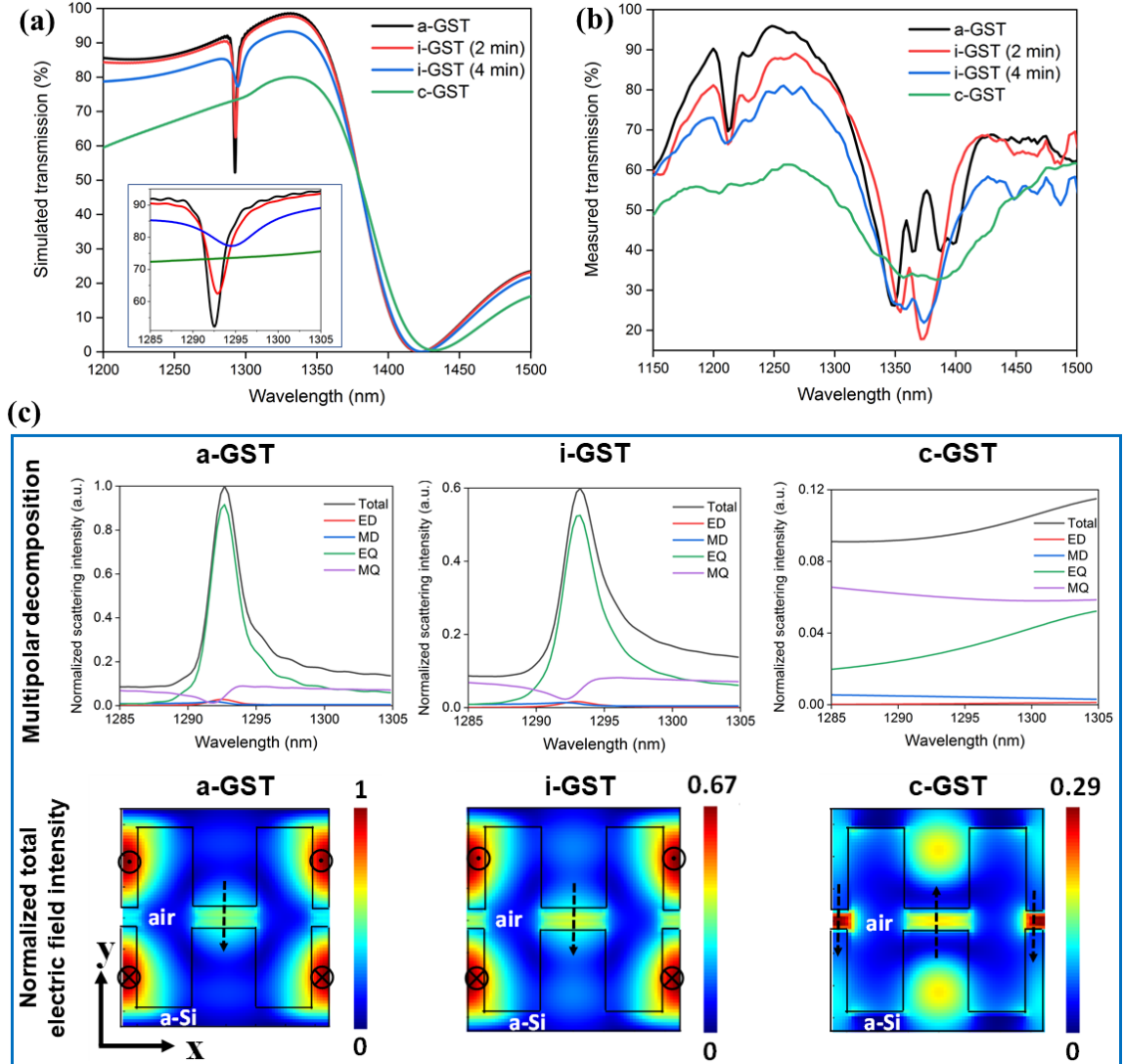

**Figure 2.** Hybrid silicon-GST nanostructure with tunable Fano-resonance. (a) Simulated transmission spectra of amorphous, intermediate states (annealing at 180 °C for 2 min and 4 min, respectively) and crystalline GST. (b) The corresponding measured transmission spectra for different states. (c) Top row: Multipolar decomposition of one unit-cell of the proposed hybrid metasurfaces for amorphous state (a-GST), intermediate state (2 min annealing time) (i-GST) and crystalline state (c-GST). Bottom row: simulated normalized total electric field intensity for the cases of a-GST, i-GST (2 min) and c-GST. The out-of-plane arrows refer to oscillation of electric quadrupole (*EQ*), and in-plane arrows indicate electric dipole (*ED*) oscillations. For each sub-figure of the field profile, it refers to the horizontal plane at the position of $z=200$ nm.

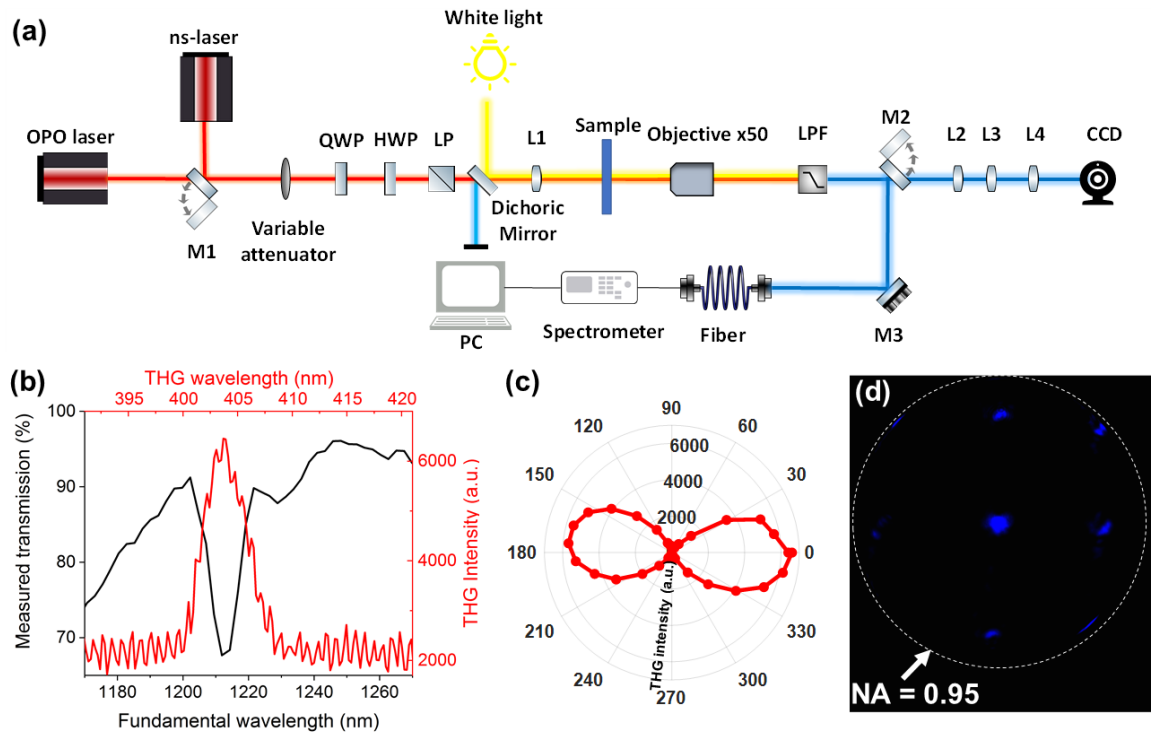

**Figure 3.** Optical characterization of THG emission at a-GST state. (a) The experimental optical setup for transmission measurement of the proposed metasurfaces using ns-laser, THG power and spectrum using optical parametric oscillator laser, and back focal plane imaging of THG using 4f-telescope setup. (b) The measured transmission spectrum of the hybrid metasurfaces and the corresponding THG emission. (c) The measured THG intensity versus the pump laser polarization. (d) Back focal plane imaging of the transmitted THG signal using an objective with a high numerical aperture (NA = 0.95), showing a high concentration of the THG emission at the zero-order diffraction.

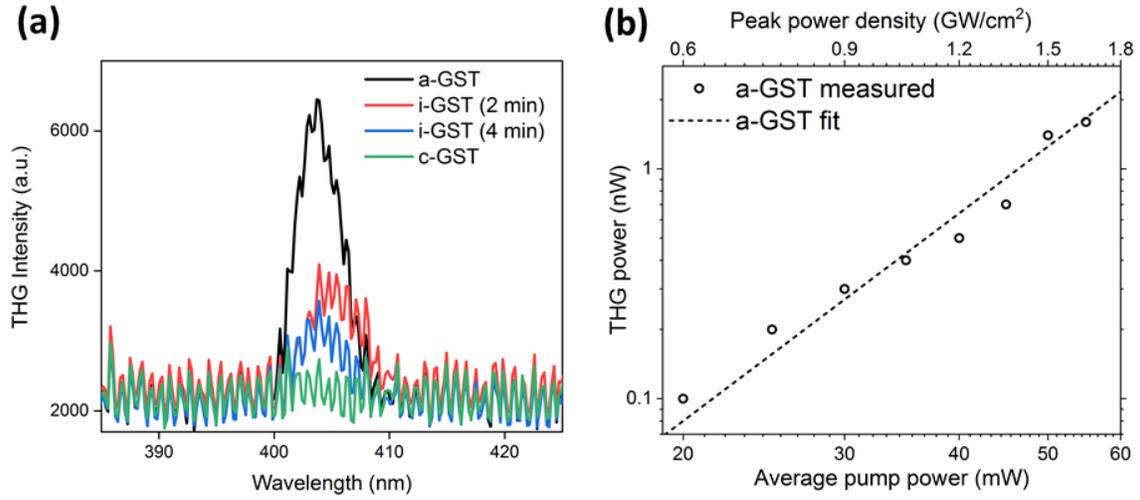

**Figure 4.** Tunable THG measurement and characterization. (a) Measured THG spectrum at different GST states. (b) The measured THG power in a log scale for a-GST state with a slope of line fit set to 3.